\documentclass[epj]{svjour}
\usepackage{graphics,color,graphicx}
\begin{document}
\definecolor{Black}{named}{Black}
\definecolor{Blue}{named}{Blue}
\definecolor{Red}{named}{Red}
\definecolor{Green}{named}{ForestGreen}
\definecolor{Black}{named}{Black}
\definecolor{Olive}{named}{OliveGreen}
\definecolor{Royal}{named}{RoyalBlue}
\definecolor{Orange}{named}{YellowOrange}
\definecolor{Yellow}{named}{Goldenrod}
\definecolor{Cornblue}{named}{CornflowerBlue}
\definecolor{Lila}{named}{DarkOrchid}
\title{Prototype for an Undulator-based Source for Polarised
Positrons} \subtitle{International Polarised Positron Collaboration:
Project E-166\thanks{{\scriptsize E-166 Collaboration:
G. Alexander, P. Anthony, V. Bharadwaj, Y. Batygin, T. Behnke,
S. Berridge, G.R. Bower, W. Bugg, R. Carr, E. Chudakov,
J.E. Clendenin, F.J. Decker, Y. Efremenko, T. Fieguth, K. Fl\"ottmann,
M. Fukuda, V. Gharibyan, T. Handler, T. Hirose, R.H. Iverson,
Y. Kamychkov, H. Kolanowski, T. Lohse, C. Lu, K. McDonald, N. Meyners,
R. Michaels, A.A. Mikhailichenko, K. M\"onig, G. Moortgat-Pick,
M. Olson, T. Omori, D. Onoprienko, N. Pavel, R. Pitthan, R. P\"oschl, 
M. Purohit,
L. Rinolfi, K.P. Sch\"uler, J.C. Sheppard, S. Spanier, A. Stahl,
Z.M. Szalata, J. Turner, D. Walz, A. Weidemann, J. Weisend}}}
\author{Gudrid Moortgat-Pick\inst{1} 
}                     
%
%
\institute{IPPP, Institute for Particle Physics 
Phenomenology, University of Durham, Durham, DH1 3LE, UK}
\date{Received: date / Revised version: date}
%
\abstract{ The full exploitation of the physics potential of a future
Linear Collider requires the development of polarised positron
beams. A very promising scheme for the technical realisation is the
use of helical undulators, generating circular polarised photons of
several MeV which are then converted in a thin target to
longitudinally polarised positrons. 
The experiment E-166 tests this scheme.
It uses the low-emittance 50-GeV electron beam at the  Final Focus Test
Beam (FFTB) at SLAC, passing through a 1 meter-long helical undulator.
The flux and polarisation of the undulator photons as well as the properties
of the positrons will be measured and will be compared with simulations.
} 
\maketitle

\vspace*{-.6cm}
\section{Introduction}
\label{intro}

\vspace*{-.4cm}
Polarised electrons have been a part of the different Linear
Collider proposals for a long time; the importance of beam
polarisation in general was demonstrated at the SLAC Linear
Collider (SLC), where during its last run 1997/98, an average
longitudinal beam polarisation $P(e^-)=74\%$ was reached.

\begin{figure}
\setlength{\unitlength}{1cm}
\resizebox{0.43\textwidth}{!}{%
  \includegraphics[width=0.5\textwidth,
height=.2\textheight]{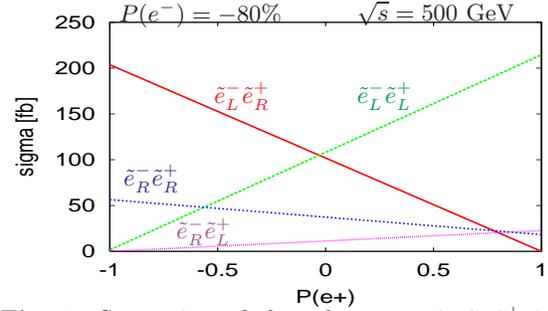}}
\put(-4.9,2.7){\small {{\color{Red}
$\tilde{e}^-_{L} \tilde{e}^+_{R}$}}}
\put(-6.1,1.6){\small{{\color{Blue}
$\tilde{e}^-_{R} \tilde{e}^+_{R}$}}}
\put(-5.4,.9){\small{{\color{Lila}
$\tilde{e}^-_{R} \tilde{e}^+_{L}$}}}
\put(-3,2.7){\small {{\color{Green}$
\tilde{e}^-_{L} \tilde{e}^+_{L}$}}}
\put(-4,3.7){\small \makebox(0,0)[br]{{$P(e^-)=-80\%$}}}
\put(-3,3.8){\small $\sqrt{s}=500$~GeV}
\vspace{-.2cm}       
\caption{Separation of the
selectron pair $\tilde{e}_L^-\tilde{e}^+_R$ in
$e^+ e^-\to \tilde{e}^+_{L,R} \tilde{e}^{-}_{L,R}$ with longitudinally
polarised beams
in order to test the association of chiral quantum numbers to scalar fermions
in SUSY transformations}
\label{fig_sel}       
\end{figure}

\begin{figure}
\setlength{\unitlength}{1cm}
\begin{picture}(10,3)
\resizebox{.4\textwidth}{!}{%
  \includegraphics[angle=90,width=0.33\textwidth,
height=.1\textheight]{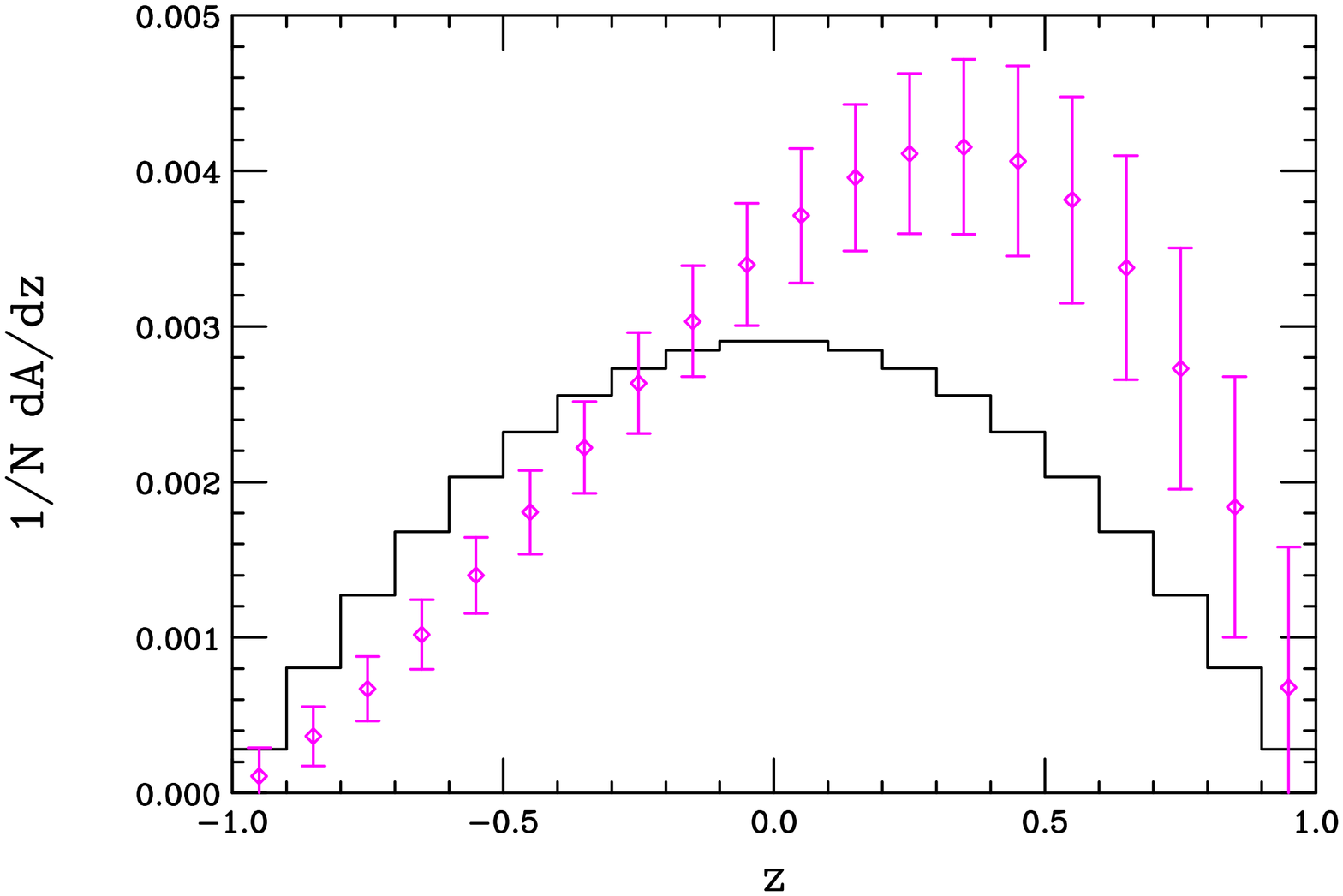}
}
\put(-2,2.9){\small $\sqrt{s}=500$~GeV}
\put(-6.6,2.9){\small $P^T(e^-)=80\%$, $P^T(e^+)=60\%$, }
\put(-5.2,2.2){\small\color{Lila} ADD model}
\put(-2.8,1.3){\small SM}
\end{picture}
\vspace{-.5cm}
\caption{Search for large extra dimensions in the ADD model in
$e^+e^-\to f\bar{f}$ with transversely polarised beams. Shown is the
differential azimuthal asymmetry distribution whose asymmetric
distribution is the signal for the graviton spin-2 exchange. 
\label{fig_ed}\vspace*{-.7cm} }
\end{figure}

Recently much scrutiny has been given to the case for polarised
positrons in addition to polarised electrons. Having both beams
polarised leads to, e.g., the well--known effect
of increasing the effective
polarisation $P_{eff}=(P(e^-)-P(e^+))/(1-P(e^-)P(e^+))$ and reducing
the relative error, see \cite{Hirose}. Moreover, it is a very
efficient tool for analysing non-standard couplings of new physics.
E.g. SUSY transformations associate chiral (anti)fermions to scalars
$e^-_{L,R}\leftrightarrow \tilde{e}^-_{L,R}$ but
$e^+_{L,R}\leftrightarrow \tilde{e}^+_{R,L}$. In order to prove this
association the use of both beam polarised is necessary
\cite{Bloechi,e166}. As can be seen in Fig.~\ref{fig_sel}, where 
the masses of the SUSY particles were chosen to be close
together, 
$m_{\tilde{e}_L}=200$~GeV, $m_{\tilde{e}_R}=190$~GeV, 
the separation of both pairs $\tilde{e}^-_{L}\tilde{e}^+_R$,
$\tilde{e}^-_{L}\tilde{e}^+_L$ is only possible with both beam
polarised. Even $P(e^-)=-100\%$ will not change the situation
substantially. Another option when polarising both beams is the use of
transversely polarised beams, where the cross section is then composed
by: $\sigma=\sigma^{\mbox{unpol+long}}+P^T(e^+) P^T(e^-)
\tilde{\sigma}^{\mbox{trans}}$.
 The use of transversely polarised beams is an efficient tool
for discovering, e.g., large extra dimension in $e^+e^-\to f \bar{f}$ and 
distinguishing different models \cite{Rizzo}.
The azimuthal asymmetry is symmetric in SM-like interactions. However, 
the exchange of the Graviton, Spin 2, particle leads to an asymmetric 
dependence, see Fig.~\ref{fig_ed}. More examples can be found 
in~\cite{pol-overview}.

\vspace*{-.5cm}
\section{Technical layout of E-166}

\begin{figure}
\setlength{\unitlength}{1cm}
\resizebox{0.5\textwidth}{!}{%
  \includegraphics{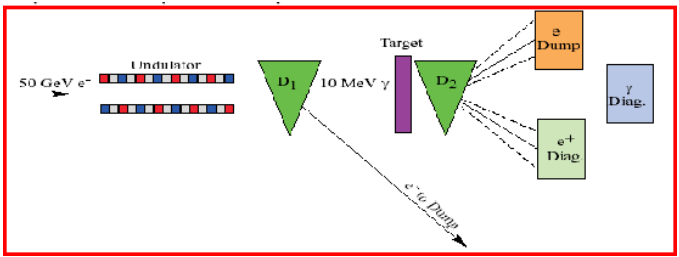}
}
\vspace{-.4cm}       
\caption{Conceptual layout of the experiment to demonstrate the
production of polarised positrons in the SLAC FFTB. The 50-GeV $e^-$
beam passes through an undulator, producing a beam of circular
polarised photons of MeV energy. The electrons are deflected by
the $D_1$ magnet. The photons are converted to electrons and positrons
in a thin Ti target.The polarisation of the positrons and photons are
measured in polarimeters based on Compton scattering of electrons in
magnetised iron.}
\label{fig-skizze}       
\end{figure}

\vspace*{-.4cm}
\begin{sloppypar}
The SLAC experiment E-166 is a demonstration of undulator-based positron
production for Linear Colliders (LC). It
employs a helical undulator \cite{Balakin}
to generate photons of
several MeV with circular polarisation which are then converted in a
thin ($\sim 0.5$ radiation length) target to generate longitudinally
polarised positrons. The experiment will install a 1-meter-long,
short-period ($\lambda_u=2.4$~mm, $K=0.17$), pulsed helical undulator
in the Final Focus Test Beam (FFTB) at SLAC, see Fig.~\ref{fig-skizze} 
\cite{e166}. 
A low-emittance 50-GeV
electron beam passing through this undulator will generate circularly
polarised photons with energies mainly up to the $1^{st}$ harmonic cutoff
energy of about 10~MeV. These polarised photons are then converted in
a $\sim 0.5$ radiation length Ti-alloy target to polarised positrons via pair
production, see Fig.~\ref{fig-skizze}. As can be seen from
Table~\ref{tab:1}, the photons produced in E-166 are in the same energy
range and with the same polarisation characteristics as for a LC.
Concerning the pair production process the same target thickness and
material as in the LC are used, however, the positron intensity/pulse
is lower by a factor 1/2000 compared to a positron source of a future LC.
\end{sloppypar}

\vspace*{-.4cm}
\subsection{Undulator design}

\vspace*{-.4cm}
The $\gamma$-rays are the result of backscattering of an electron beam
of energy $E_e$ off the virtual photon of an undulator with period
$\lambda_U$.  To create positrons, $\gamma$-rays of at least a 
few MeV
are needed.  The intensity of the $\gamma$-rays depends on the
intensity of the virtual photons, and hence on the square of its
magnetic field strength, which is measured via the dimensionless
undulator parameter $K=0.09 B_0[T]\lambda_U[mm]$.
The undulator radiation is given by
\begin{center}
${\frac{dN_{\gamma}}{dL}}=\frac{30.6}{\lambda_u[\mbox{\small mm}]}
\frac{K^2}{1+K^2} \mbox{\small photons}/m/e^-=0.37 \mbox{\small photons}/e^-
$
\end{center}
and this photon number spectrum is rather flat up to the maximum energy 
$E_{c10}$ of the first harmonic radiation
\begin{center}
$
{E_{c10}}=24[\mbox{\small MeV}]
\frac{(E_e/50[\mbox{\small GeV}])^2}{{\lambda_u}[\mbox{\small mm}](1+
K^2)}{=9.6} \mbox{\small MeV}
$
\end{center}
Since the highest practical beam energy at SLAC is 50 GeV, one chooses
$\lambda_U=2.4$~mm and $K=0.17$.

The helical undulator is 1-m long, consists of a 0.6-mm-diameter
copper wire bifilar helix, wound on a stainless-steel support tube,
whose inner diameter is 0.889 mm. The on-axis field in the undulator
is 0.76 T for a 2300-A excitation. The undulator is immersed in an oil
bath for cooling.

\vspace*{-.4cm}
\subsection{Production of polarised positrons}

\begin{figure*}[t]
\setlength{\unitlength}{1cm}
\hspace{-.5cm}
\begin{minipage}{18cm}
  \includegraphics[height=.275\textheight,
width=0.33\textwidth]{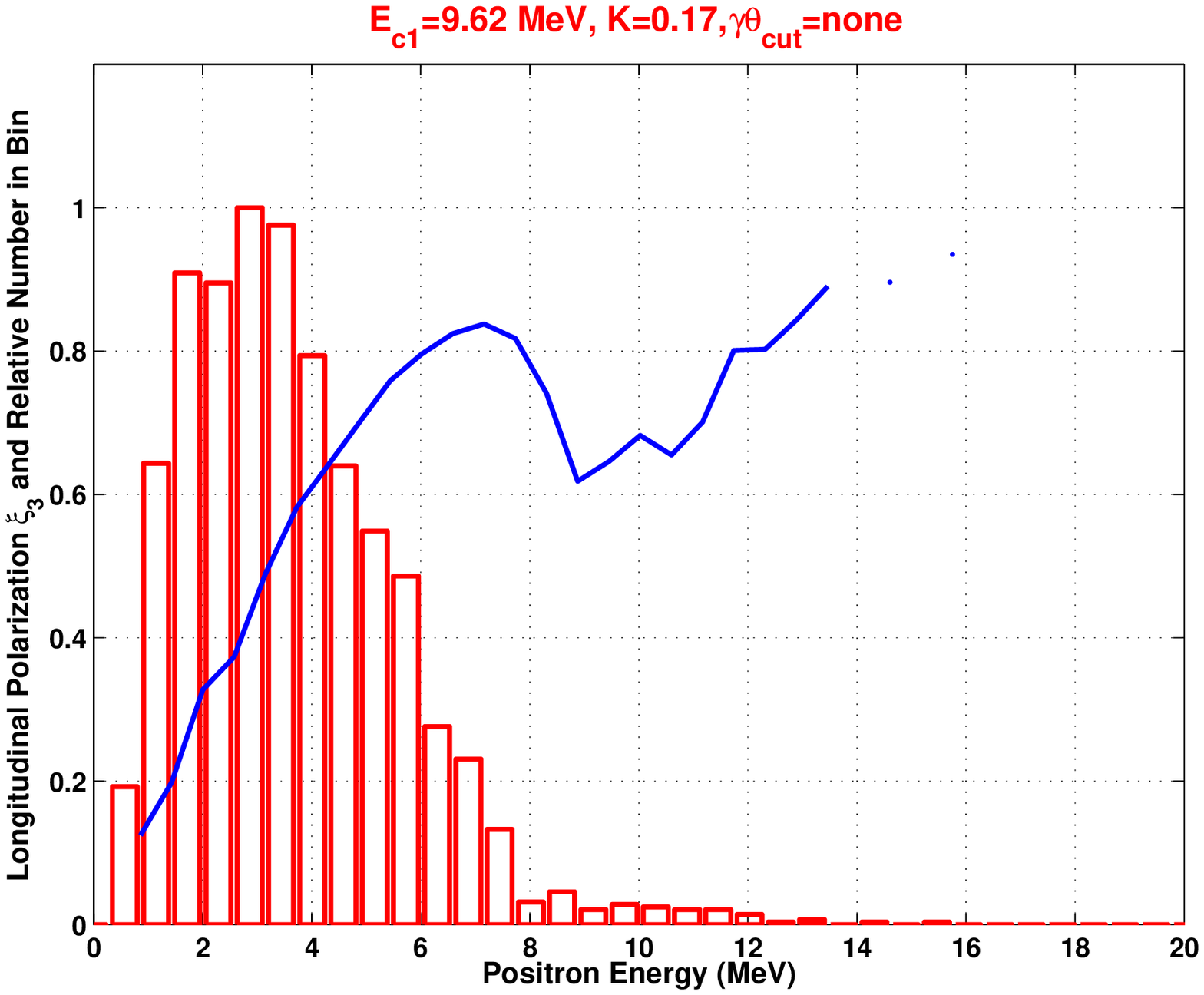}
\end{minipage}
\hspace*{5.8cm}
\begin{minipage}{18cm}
\vspace{-6.8cm}
\resizebox{.7\textwidth}{!}{%
  \includegraphics{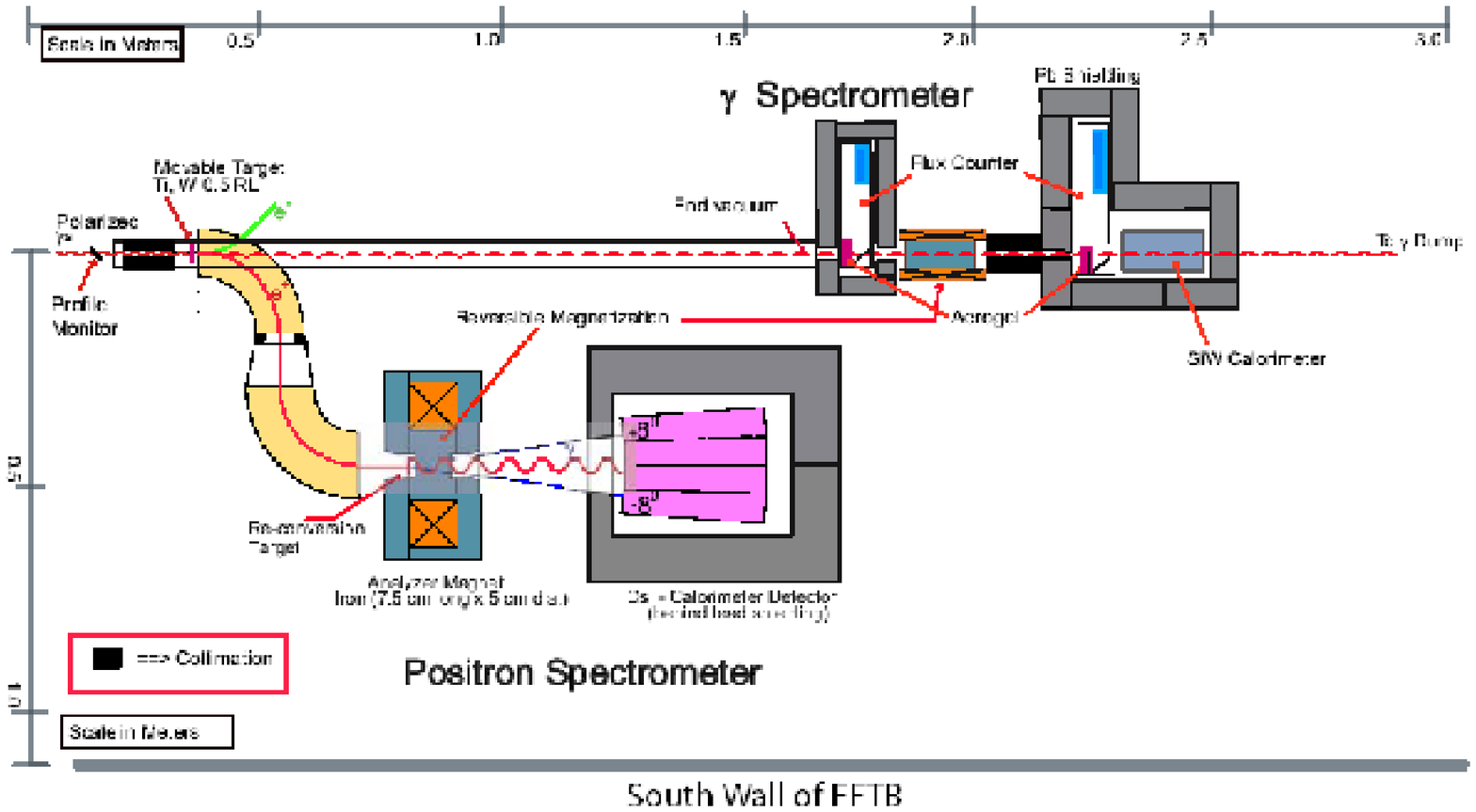}
}
\end{minipage}
\vspace{-.5cm}       
\caption{{\bf Left:} Longitudinal
polarisation (solid) and energy spectrum (histogram) of positrons
emitted for the chosen helical undulator design;
{\bf Right:} Conceptual layout of the E-166 positron generation and
photon and positron diagnostic systems}
\label{fig_spek}       
\end{figure*}

\vspace*{-.4cm}
The polarisation state of the photon is transferred to the outgoing
electron-positron pair in a thin target according to the cross section
derived in \cite{Olsen}.  Positrons with an energy close to the energy
of the incoming photons are 100\% longitudinally polarised, while
positrons with a lower energy have a lower polarisation. Due to an
interplay between energy loss via bremsstrahlung followed by a slight
loss of polarisation,   
the polarisation of positrons of a given energy is maximal in
targets of up to 0.5 radiation length. In the E166 undulator design
the positrons are generated at a 0.5 radiation-length-thick Titanium
target, with a longitudinal polarisation and energy spectrum as
shown in Fig.~\ref{fig_spek} (left). The composite polarisation of the
total sample is about 53\%.

\vspace*{-.4cm}
\subsection{Polarimetry at E-166}
\label{sec:1}

\vspace*{-.4cm}
The measurement of the circular polarisation of energetic photons are
based on the spin dependence of Compton scattering off atomic
electrons. In E166 the transmission of unscattered photons through a
thick magnetised iron absorber is used for the MeV $\gamma$-ray
polarimetry \cite{polarimetry}. The spin dependent part of the Compton
scattering cross section is given by $\sim P_{\gamma} P_e \sigma_P$,
where $P_{\gamma}$ is the net polarisation of the photons, $P_e$ the
net polarisation of the atomic electrons ($\pm 7.92\%$ for iron
saturation) and $\sigma_P$ is the polarised Compton scattered 
cross section. The spin dependent part of the transmission probability 
is given by
\begin{center}
$
T^{\pm}(L)\sim exp[\pm n L P_e P_{\gamma} \sigma_P],
$
\end{center}
where $n$ is the number density of atoms in iron and $L$ the length of
the iron. It turns out that for about 7.5-MeV photons a 15-cm-thick
magnetised iron absorber will become optimal, minimising the
photon background radiation in the detector.

The photon polarimeter, see Fig.~\ref{fig_spek} (right) includes two
types of detectors, a total absorption SiW calorimeter and an aerogel
Cerenkov detector; the latter one is only sensitive to photons with an energy
above 5 MeV and therefore independent of possible backgrounds of
lower-energy photons.

The positron polarimeter consists of a 2-step process: the
reconversion of the positrons at a 0.5-rad-length-thick Titanium target into
polarised photons and the polarisation measurement (again via
transmission polarimetry) of the obtained photons (typical energy of about 
$\sim 1$~MeV) with a CsI detector.

Geant simulations have shown that systematic errors 
of maximal up to $\Delta(P)/P\sim 5\%$ are expected. This confirms that
transmission polarimetry is well suited for E-166.
More details about the technical layout of E-166 can be found in \cite{e166}.

\vspace*{-.4cm}
\subsection{Outlook}
\label{sec:2}

\vspace*{-.4cm}
The project E-166 was approved in June 2003,
and will be scheduled for several weeks of running time in
January 2005 at the FFTB at SLAC.\\

\begin{sloppypar}
\noindent The author would like to thank the organisers of the EPS2003 at 
Aachen
for the wonderful and very interesting conference!
GMP would like to thank all E-166 members for the interesting and 
constructive
collaboration and the many, with great pleasure expected, weekly 
conference calls.
\end{sloppypar}

\vspace*{.8cm}

\vspace*{-1cm}
\vspace*{-.5cm}

%
%
\begin{table}
\caption{TESLA, NLC, E-166 polarised positron parameters}
\label{tab:1}       
\hspace{-.5cm}
{\small
\begin{tabular}{|lccc|}
\hline\noalign{\smallskip}
Parameter& TESLA & NLC & E-166\\
\noalign{\smallskip}\hline\noalign{\smallskip}
Beam Energy, $E_e$ [GeV] & 150-250 & 150 & 50\\
$N_e$/bunch & $3\times 10^{10}$ & $8\times 10^9$ & $1\times 10^{10}$\\
$N_{bunch}$/pulse & 2820 & 190 & 1\\
Pulses/s [Hz] & 5 & 120 & 30\\
Undulator Type & plan./helical & helical & helical\\
Und. Parameter, $K$ & 1 & 1 & 0.17\\
Und. Period, $\lambda_u$ [cm] & 1.4 & 1.0 & 0.24\\
Und. Length, $L$ [m] & 135 & 132 & 1\\
$1^{st}$ Harmon., $E_{c10}$ [MeV] & {9-25}&{11}&
{9.6}\\
$\mbox{d}N_{\gamma}/\mbox{d}L$ [$\gamma$/m/$e^-$] & 1 & 2.6 & 0.37\\
Target Material & {Ti-alloy} & {Ti-alloy} &
{Ti-alloy,}\\
&&& W\\
Target Thickn. [rad. len.] & {0.4} & {0.5} &
{0.5}\\
\noalign{\smallskip}\hline
\end{tabular}
}
\vspace*{-.2cm}  
\end{table}

\begin{table}
\hspace{1cm}
\caption{Parameters of the helical undulator system}
\label{tab:2}
\hspace{.5cm}
{\small
\begin{tabular}{|lcc|}
\hline\noalign{\smallskip}
Parameter & Units & Value\\
\noalign{\smallskip}\hline\noalign{\smallskip}
Length  & m & 1.0\\
Inner Diameter & mm & 0.89\\
Period & mm & 2.4\\
Field & kG & 7.6\\
Undulator Parameter, K & -- & 0.17\\
Current & Amps & 2300\\
Peak Voltage & Volts & 540\\
Pulse Width & $\mu$s & 30\\
Inductance & H & 0.9$\times 10^{-6}$\\
Wire Type & --& Cu\\
Wire Diameter& mm & 0.6\\
Resistance & ohms & 0.110\\
Repetition Rate & Hz& 30\\
Power Dissipation & W & 260\\
$\Delta T$/pulse & C & 2.7\\
\noalign{\smallskip}\hline
\end{tabular}
}
\vspace*{-.4cm} 
\end{table}


\begin{thebibliography}{}
\bibitem{Hirose} \vspace*{-.2cm}
T. Omori, {\it A Polarized Positron Beam for Linear 
Colliders}, KEK 98-237;
T. Hirose et al., 
JLC, Nucl. Instr. and Meth. {\bf A455}, 15 (2000).
\bibitem{Bloechi} C.~Blochinger, H.~Fraas, G.~Moortgat-Pick, W.~Porod,
Eur.\ Phys.\ J.\ C {\bf 24} (2002) 297
[hep-ph/0201282].
\bibitem{e166} G. Alexander et al, {\it Undulator-Based Production of 
Polarized Positrons}, SLAC-PROPOSAL-E-166,
LC-DET-2003-044; http://www.slac.stanford.edu/exp/e166.
\bibitem{Rizzo} T.G.~Rizzo, JHEP {\bf 0302}, 008 (2003) [hep-ph/0211374].
\bibitem{pol-overview}
G.~Moortgat-Pick, H.~M.~Steiner,
Eur.\ Phys.\ J.\ directC {\bf 3} (2001) 6 [hep-ph/0106155];
 G.~Moortgat-Pick, hep-ph/0303234; see also webpage of the polarisation 
working group `POWER': http://www.ippp.dur.ac.uk/$\tilde{}$ gudrid/power.
\bibitem{Balakin} V.E. Balakin, A.A. Mikhailichenko, {\it The
Conversion System for Obtaining High Polarized Electrons and
Positrons}, Budker Institute of nuclear Physics, Preprint BINP 79-85
(1979).
\bibitem{Olsen} H. Olsen, L.C. Maximon, 
Phys. Rev. {\bf 114}, 887 (1959).
\bibitem{polarimetry}
H. Schopper, Nucl. Instr. and Meth. {\bf 3}, 158 (1958);
M. Fukuda, T. Aoki, K. Dobashi, T. Hirose, T. Iimura, Y. Kurihaya, T. Okugi,
T. Omori, I. Sakai, J. Urakawa, M. Washio, Phys. Rev. Lett. {\bf 91} (2003).
\end{thebibliography}
\end{document}